\documentclass[12pt]{article}

\makeatletter
\@ifundefined{date}{}{\date{}}

\usepackage{times}

\usepackage{amsmath,amssymb,enumerate}
\usepackage[bf]{caption2}

\usepackage{epsfig}
\usepackage{bbm}

\title{Shaping frequency entangled qu$d$its}

\author
{Christof Bernhard$^{\ast}$, B\"anz Bessire$^{\ast}$, Thomas Feurer, Andr\'e Stefanov$^{\dagger}$\\
\\
\normalsize{Institute of Applied Physics, University of Bern, 3012 Bern, Switzerland.}
\\
\normalsize{$^\dagger$To whom correspondence should be addressed. E-mail:  andre.stefanov@iap.unibe.ch.}\\
\normalsize{$^{\ast}$These authors contributed equally to this work.}
}

\makeatother

\begin{document}

\baselineskip24pt

\maketitle

\begin{abstract}
Quantum entanglement between qu$d$its - the $d$-dimensional version of qubits - is relevant for advanced quantum information processing and provides deeper insights in the nature of quantum correlations. Encoding qu$d$its in the frequency modes of photon pairs produced by continuous parametric down-conversion enables access to high-dimensional states. By shaping the energy spectrum of entangled photons, we demonstrate the creation, characterization and manipulation of entangled qu$d$its with dimension up to 4. Their respective density matrices are reconstructed by quantum state tomography. For qubits and qutrits we additionally measured the dependency of a $d$-dimensional Bell parameter for various degrees of entanglement. Our experiment demonstrates the ability to investigate the physics of high-dimensional frequency entangled qu$d$it states which are of great importance for quantum information science. 
\end{abstract}

\section{Introduction}

Entanglement is one of the most intriguing features of quantum theory \cite{horodecki2009}. It has been experimentally revealed by the observation of correlations with no classical origins. Specifically, through Bell inequalities, the non-locality of nature has been tested by numerous experiments using entangled two-dimensional states (qubits) \cite{zeilinger1999}. Entangling $d$-dimensional states, denoted as qu$d$its, allows to formulate generalized Bell inequalities, which are more resistant to noise than their two-dimensional predecessors and lowers the threshold of the detection efficiency for loophole free Bell experiments \cite{collins2002,kaszlikowksi2002,acin2002,vertesi2010}. Entanglement is also a fundamental resource for quantum information. Here, entanglement in higher dimensions is a prerequisite for implementing more complex protocols. For instance, the effective bit rate of quantum key distribution (QKD) can be enhanced, while at the same time the secret key rate and the robustness to errors increases with $d$ \cite{sheridan2010}.\\
Due to their low decoherence rate, photons are used in many experiments as a robust carrier of entanglement. Photonic entangled states are usually produced by the nonlinear interaction of spontaneous parametric down-conversion (SPDC) \cite{klyshko1988}. The coherence of this process, together with conservation rules, can generate entanglement in the finite Hilbert space of polarization states \cite{kwiat1999}. Entanglement in infinite Hilbert spaces can be realized for transverse (momentum) or orbital angular momentum (OAM) modes \cite{pires2009,mair2001,dada2011,agnew2011,fickler2012,giovannini2012} and for energy-time states \cite{law2000}. The amount of entanglement is commonly quantified by the Schmidt number $K$. For transverse wave vector entanglement $K$ is on the order of 10 for perfect SPDC phase matching conditions \cite{law2004}, approximately 400 for specific non-perfect phase matching conditions \cite{pires2009}, and approximately 50 for OAM entanglement \cite{giovannini2012}.\\
Similar Schmidt numbers can be achieved in continuous energy-time entanglement generated by short pump pulses \cite{law2000,mikhailova2008} but much larger $K$ numbers are obtained for a quasi-monochromatic pump laser. In order to access these high-dimensional states, an experimental setup with many control parameters is required. In practice, the infinite Hilbert space is projected onto a finite space of, for example, discrete time- or frequency-bins. In the time-bin subspace with $d=3,4$ two-photon interferences have been observed by interferometers with multiple arms \cite{thew2004,richart2012}. This, however, requires interferometric stability and becomes prohibitively complex for higher dimensions. In the frequency-bin subspace, interferences between two entangled photons, each in an effective two-dimensional space, have been observed by manipulating the spectra with a combination of narrowband filters and electro-optic modulators \cite{olislager2010}. In this approach a complex modulation scheme would be needed to address qu$d$its in higher dimensions. Besides being very useful for QKD implementations, the aforementioned experiments do not provide sufficient control of the phase and amplitude of qu$d$its in the frequency domain to extensively study the properties of $d$-dimensional states with $d>2$.\\
Here, we demonstrate a methodology that allows for full control over entangled qu$d$its through coherent modulation of the photon spectra. It is derived from a classical pulse shaping arrangement and contains a spatial light modulator (SLM) as a reconfigurable modulation tool. This technique is widely used in ultrafast optics \cite{weiner2000} and has been adapted to manipulate the wavefunction of energy-time entangled two-photon states \cite{peer2005,zaeh2008}. The flexibility of the experimental setup enables the generation, characterization, and manipulation of $d$-dimensional qu$d$it states. We realize qu$d$its with dimensions of up to 4. Specifically, we verify the generation of the maximally entangled qu$d$it states through tomographic quantum state reconstruction. Subsequent Bell-type measurements demonstrate the applicability of the reconstructed quantum states. The versatility of the SLM based setup allows to test theoretical predictions beyond two-dimensional entangled states. As a first demonstration, we investigate the sensitivity of the Bell parameter for maximally and non-maximally entangled qubit and qutrit states by varying the degree of entanglement.\\

\section{Discretization of frequency-entangled photons}

We consider entangled photons generated in a SPDC process of type-0 where all involved photons, the pump, the created idler ($i$) and signal ($s$) photon are identically polarized \cite{lerch2013}. Restricting the configuration of the three photons to the case where they are mutually collinear, the entangled two-photon state can then be written as \begin{equation}
\vert\psi\rangle = \int_{-\infty}^{\infty}\int_{-\infty}^{\infty} d\omega_id\omega_s \Lambda(\omega_i, \omega_s)\hat{a}^{\dagger}_{i}(\omega_i)\hat{a}^{\dagger}_{s}(\omega_s)\vert0\rangle_i\vert0\rangle_s
\end{equation}
where $\Lambda(\omega_i,\omega_s)=\alpha(\omega_i,\omega_s)\Phi(\omega_i,\omega_s)$ describes the joint spectral amplitude of SPDC in terms of the pump envelope function $\alpha(\omega_i,\omega_s)$ and the phase matching function $\Phi(\omega_i,\omega_s)$. Idler and signal photons with corresponding relative frequency $\omega_{i,s}$ are created by the operators $\hat{a}^{\dagger}_{i,s}(\omega_{i,s})$, acting on the combined vacuum state \mbox{$\vert0\rangle_i\vert0\rangle_s$}. We have calculated the entropy of entanglement $E=-\mbox{Tr}(\hat{\rho}\log_2\hat{\rho})$ of $\Lambda(\omega_i,\omega_s)$ to be $E=(22.0\pm 0.2)$ ebits for a pump bandwidth of 5 MHz by means of a numerical approximation method \cite{bennett1996,wihler2012}. Here, $\hat{\rho}$ denotes the reduced density operator of the idler or signal photon subsystem. This amount of entropy is the same as in a maximally entangled qu$d$it state with $d=2^{E}\approx 4.2\cdot 10^6$. As a further quantification of entanglement, the Schmidt number $K=1/\mbox{Tr}(\hat{\rho}^2)$ has been computed numerically to be $K\approx 2.4\cdot 10^6$. In order to use this large resource of entanglement for quantum information processing, we encode qu$d$its in the frequency domain by projecting the state $\vert\psi\rangle$ into a discrete $d^2$-dimensional subspace spanned by the states $\vert j \rangle_i \vert k\rangle_s$ with $\left\vert j\right\rangle_{i,s} \equiv\int_{-\infty}^{\infty}d\omega f^{i,s}_j\left(\omega\right)\hat{a}^{\dagger}_{i,s}(\omega)\left\vert 0\right\rangle_{i,s}$ and $j=0,...,d-1$.
The projected state is then expressed by
 \begin{equation}\label{eq:psidisc}
\vert\psi\rangle^{(d)} = \sum_{j=0}^{d-1}\sum_{k=0}^{d-1} c_{jk}\vert j \rangle_i \vert k\rangle_s
\end{equation}
with coefficients $c_{jk}  = \int_{-\infty}^{\infty}\int_{-\infty}^{\infty}d\omega_{i}d\omega_{s}f_{j}^{i*}\left(\omega_{i}\right) f_{k}^{s*}\left(\omega_{s}\right)\Lambda\left(\omega_{i},\omega_{s}\right)$. The functions $ f^{i,s}_j\left(\omega\right)$ can be  chosen arbitrarily under the condition to be orthogonal $\int_{-\infty}^{\infty} d\omega f^{i,s \ast}_j(\omega)f^{i,s}_k(\omega) = \delta_{jk}$. For the experiments presented here, we specifically define frequency-bins according to
\begin{equation}\label{eq:fbins}
f_{j}^{i,s}(\omega)=\begin{cases}1/\sqrt{\Delta\omega_j} &\text{for}\left|\omega-\omega_{j}\right|<\Delta\omega_j \\
0&\text{otherwise.}\end{cases} 
\end{equation}
Imposing $|\omega_j-\omega_k|>(\Delta\omega_j+\Delta\omega_k)/2$ for all $j,k$ ensures that adjacent bins do not overlap. For simplicity, we assume in the following a continuous wave pump by $\alpha(\omega_i, \omega_s)\propto\delta(\omega_i+\omega_s)$ and therefore $\vert\psi\rangle^{(d)}$ becomes restricted to its diagonal form
\begin{equation}\label{eq:psidiag}
\vert\psi\rangle^{(d)} = \sum_{j=0}^{d-1} c_{j}\vert j \rangle_i \vert j\rangle_s.
\end{equation}\\

\section{Experimental setup}
\begin{figure}[ht!]
	\begin{center}
	\epsfig{file=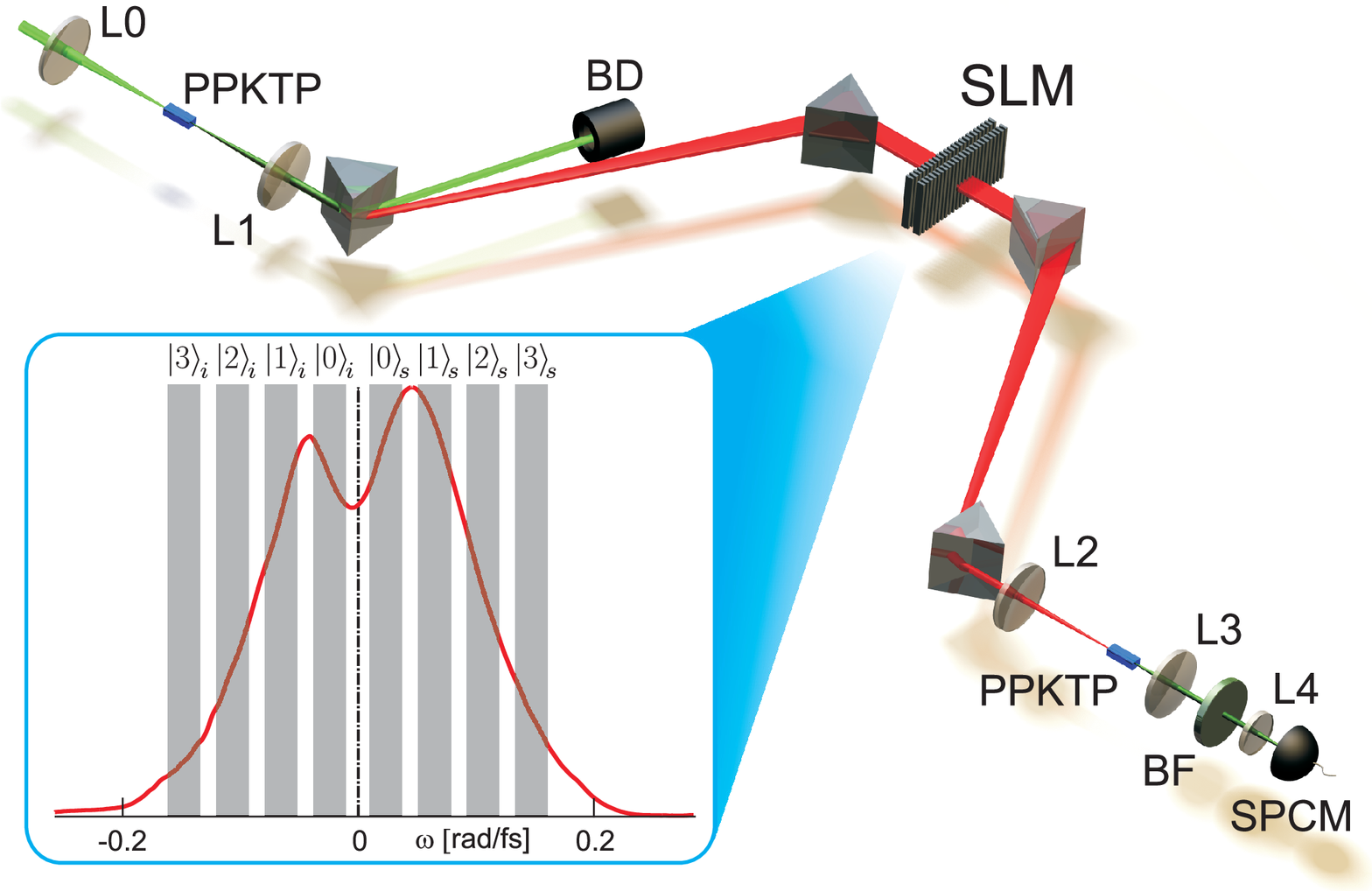,width=0.8\columnwidth}
\caption{\label{fig:Setup} Schematic of the experimental setup. L0 pump beam focusing lens (f = 150 mm), PPKTP periodically poled potassium titanium oxide phosphate crystal 1x2x11.5 mm$^3$, BD beam dump, SLM spatial light modulator, L1 and L2 two lens symmetric imaging arrangement (f = 100 mm) to enhance the spectral resolution with a magnification of 1:6, prism compressor composed of four N-SF11 equilateral prisms, BF bandpass filter 4 mm BG18 glass, SPCM single photon counting module with a two lens (L3, L4) imaging system. The inset shows the measured down-converted spectrum overlaid with a schematic illustration of the frequency-bins for a ququart. Each of the gray shaded areas represents a single bin whose amplitude and phase can be manipulated individually.}  
	\end{center}
\end{figure}

To experimentally generate type-0 entangled photons degenerated at 1064 nm, we pump a 11.5 mm long positive uniaxial and periodically poled nonlinear KTiOPO$_{4}$ (PPKTP) crystal with a poling periodicity of 9 $\mu$m by means of a quasi-monochromatic Nd:YVO$_{4}$ (Verdi) laser centered at 532 nm featuring a narrow spectral bandwidth of about 5 MHz (Fig.~\ref{fig:Setup}). The collinear pump beam is focused into the middle of the PPKTP crystal with a power of 5 W. Mounted in a temperature stabilized copper block at 30.7 $^{\circ}$C, the down-conversion crystal creates entangled photons with a spectral bandwidth of 105 nm. The measured photon flux power is 0.8 $\mu$W and corresponds to a spectral mode density of 0.15. This ensures that we are below the single photon limit and, therefore, the entangled photon pairs are temporally separated from each other \cite{dayan2005}. 
To compensate for group-velocity dispersion in the setup and to allow for coherent shaping of their spectra, idler and signal photon are imaged through a four-prism compressor arrangement, where the first prism deflects the residue of the pump into a beam dump. At the symmetry axis between the second and the third prism a SLM (Jenoptik, SLM-S640d) is aligned along the spatially dispersed down-converted spectrum. This device consists of two similar nematic liquid crystal arrays of 640 pixels, each with a width of 100 $\mu$m and separated by a gap of 3 $\mu$m. The orientation of the liquid crystal molecules within a pixel can be controlled by the applied voltage. Together with a linearly polarized input beam and a polarization dependent detection scheme, the phase and amplitude of the transmitted frequencies at each pixel can be modulated \cite{weiner2000}. The effect of the SLM on each photon is described by a complex transfer function $M^{i,s}(\omega)$. A frequency-bin structure according to Eq.~(\ref{eq:fbins}) is then implemented on the SLM through
\begin{equation}\label{eq:mslm}
M^{i,s}(\omega)=\sum_{j=0}^{d-1}u_j^{i,s}f_{j}^{i,s}(\omega)=\sum_{j=0}^{d-1}|u_j^{i,s}|e^{i\phi_j^{i,s}}f_{j}^{i,s}(\omega),
\end{equation}
where $|u_j^{i,s}|$ and $\phi_j^{i,s}$ are controlled independently. Since in our experiment there is no spatial separation between idler and signal modes, we address each photon individually by assigning $M^i(\omega)$ to the lower frequency part and $M^s(\omega)$ to the higher frequency part of the spectrum. Coincidences of the entangled photon pairs are detected within a time window of about 100 fs through up-conversion in a second PPKTP crystal \cite{peer2005}. The detection crystal is temperature stabilized at 35 $^{\circ}$C to maximize the up-conversion rate. The recombined 532 nm photons are then imaged onto the active area of a single photon counting module (SPCM, ID Quantique, id100-50-uln). Since the entangled photons are detected in coincidences through up-conversion, we define the measured state $\vert\psi\rangle_{S} = \int_{-\infty}^{\infty} d\omega \Gamma(\omega)\hat{a}^{\dagger}_{i}(\omega)\hat{a}^{\dagger}_{s}(-\omega)\vert0\rangle_i\vert0\rangle_s$ where $\Gamma(\omega)$ describes the joint spectral amplitude of SPDC filtered by the phase matching properties of the detection crystal. The measured signal after shaping and the up-conversion stage then reads $S=\left|\int_{-\infty}^{\infty}d\omega\Gamma(\omega)M^i(\omega)M^s(-\omega)\right|^2$ for a continuous wave pump and is equivalent to 
\begin{equation}\label{eq:projection}
S=\left\vert\langle\chi\vert\psi\rangle^{(d)}\right\vert^2=\left\vert\sum_{l=0}^{d-1}u^i_l u^s_l c_l\right\vert^2 
\end{equation}
for the direct product state $\vert\chi\rangle=\left(\sum_{j=0}^{d-1}u^{i*}_j\vert j\rangle_i\right)\left(\sum_{j'=0}^{d-1}u^{s*}_{j'}\vert j'\rangle_s\right)$ with $\vert\psi\rangle^{(d)}$ of Eq.~(\ref{eq:psidiag}). 
The combination of the SLM together with an up-conversion coincidence detection therefore realizes a projective measurement. Different quantum protocols can thus be implemented by the corresponding choice of $\vert\chi\rangle$. \\

\section{Quantum state tomography of maximally entangled qu$d$its}
At first, maximally entangled states are generated by Procrustean filtering \cite{bennett1996} where we equate the $c_j$ coefficients in Eq.~(\ref{eq:psidiag}) by adjusting the amplitudes $|u_j^{i,s}|$ of Eq.~(\ref{eq:mslm}). The Procrustean filtering applies in two steps. First, we place the frequency-bins of Eq.~(\ref{eq:fbins}) such that the states $\vert j\rangle_i\vert j'\rangle_s$ are well separated on the frequency axis and that no coincidence events are detected when $j \neq j'$. By doing so, we implement the orthogonality of the frequency-bins. Second, considering the down-converted spectrum (Fig.~\ref{fig:Setup}), it is obvious that the shape of the spectrum in combination with the bin width $\Delta \omega_{j}$ and its central frequency $\omega_j$ defines the amount of coincidences for the states with $j=j'$. It is then the $\vert j\rangle_i\vert j\rangle_s$ state with the lowest coincidence rate which defines in Eq.~(\ref{eq:mslm}) the amplitude scaling $|u_j^{i,s}|$ of the others. We measure net count rates of 43 Hz for a qubit, 13 Hz for a qutrit, and 6 Hz for a ququart. In an iterative procedure of scaling the amplitudes followed by measuring the $\vert j\rangle_i\vert j\rangle_s$ states, we balance the coincidence rates until they become equal within their errors assuming Poisson statistic. This filtering is a trade-off between maximal count rates and purity of the maximally entangled state and leads to the desired equation of the $c_j$ in Eq.~(\ref{eq:psidiag}). 

\begin{figure}[ht!]
\epsfig{file=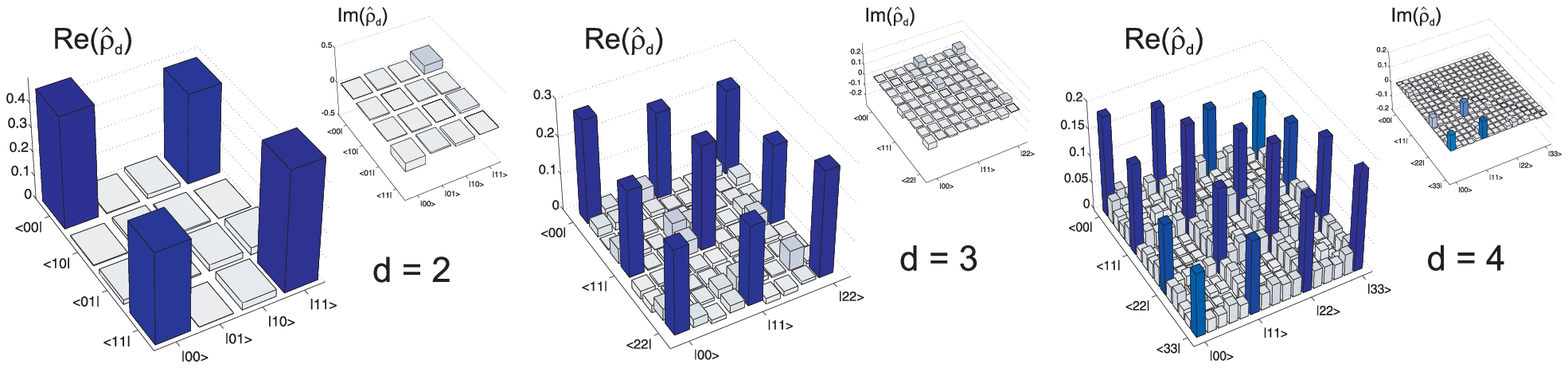,width=1\columnwidth}
\caption{\label{fig:QuditsTomo} From left to right: Reconstructed density matrices $\hat{\rho}_d$ of a qubit, qutrit, and a ququart, based on background-subtracted coincidence counts. Shown are the real and the imaginary parts. The small, residual imaginary values are due to remaining dispersion between the frequency-bins.} 
\end{figure}

Quantum-state tomography then allows us to retrieve the density matrix $\hat{\rho}_d$ of these states by performing projective measurements \cite{james2001,thew2002}. As a tomographically complete set of basis vectors we choose, inspired by \cite{agnew2011}, single bin states $\vert\chi_{j_1}\rangle = \vert j_1\rangle_i \vert j_1'\rangle_s$ and superposition of two bin states $\vert\chi_{\alpha,j_1,j_2}\rangle = \frac{1}{2}(\vert j_1\rangle_i +e^{i\alpha}\vert j_2\rangle_i)(\vert j'_1\rangle_s +e^{i\alpha'}\vert j'_2\rangle_s)$ with $j_1^{(\prime)} ,j_2^{(\prime)} = 0,...,d-1$ and $j_1^{(\prime)}<j_2^{(\prime)}$ with the relative phase $\alpha^{(\prime)} = 0, \frac{\pi}{2}$. To obtain a positive semidefinite $\hat{\rho}_d$  we employ a maximum likelihood estimation \cite{hradil1997}, first applied to a qubit in \cite{james2001} and then extended to qu$d$its up to $d=8$ in \cite{agnew2011}. The reconstructed density matrices up to dimension $d=4$ are shown in Figure \ref{fig:QuditsTomo}. As a measure of how close the reconstructed state $\hat{\rho}_d$ is to a maximally entangled state $\hat{\rho}_{me}$, we computed the fidelity $F_d = \mbox{Tr}(\sqrt{\sqrt{\hat{\rho}_{me}}\hat{\rho}_d \sqrt{\hat{\rho}_{me}}})$  \cite{jozsa1994}. When $\hat{\rho}_d$ is a maximally entangled state, the fidelity is equal to one. The computed fidelities of the reconstructed states are $F_2 = 0.928 \pm 0.010$, $F_3 = 0.855 \pm 0.010$, and $F_4 = 0.781 \pm 0.018$. We estimated the statistical 2$\sigma$-error with a  Monte-Carlo method by randomly adding normally distributed count rate errors to each measurement outcome and recomputing the fidelity. With increasing dimension we have to implement more frequency-bins within the same spectral range. Because of the finite spectral resolution of the setup, the increasing overlap between adjacent bins leads to a decrease of the fidelity.
\\

\section{Bell inequalities for non-maximally entangled qu$d$its}
More generally, we obtain non-maximally entangled states by varying the amplitudes in Eq.~\ref{eq:psidiag}. These states are then studied with regard to their non-local properties.
The non-local properties of quantum correlations are usually measured by a Bell parameter $I$ whose value is restricted under the local variables assumption. In \cite{collins2002}, Collins \textit{et al.}~(CGLMP) generalized Bell inequalities to  arbitrary $d$-dimensional bipartite quantum systems by defining a dimensional dependent Bell parameter $I_d$. If correlations between two space-like separated systems can be explained through local realism, then $I_d\leq 2$ holds for all $d\geq 2$. Despite a left-open locality loophole in our detection method, the violation of the precedent inequality indicates the existence of non-classical correlations due to entanglement. A counterintuitive property of the CGLMP inequality is that for dimensions $d\geq 3$ the inequality is more strongly violated by certain non-maximally entangled states than by maximally entangled states \cite{kaszlikowksi2002,acin2002}. In order to compare the sensitivity of the Bell parameters $I_{2}(\gamma)$ and $I_{3}(\gamma)$ to an entanglement parameter $\gamma\in[0,1]$, we consider the following bipartite qubit and qutrit states
\begin{equation}\label{eq:gamma_2}
\vert\psi(\gamma)\rangle^{(2)}=\frac{1}{\sqrt{1+\gamma^2}}(\vert 0 \rangle_A \vert 0\rangle_B+\gamma\vert 1 \rangle_A \vert 1\rangle_B),
\end{equation}
\begin{equation}\label{eq:gamma_3}
\vert\psi(\gamma)\rangle^{(3)}=\frac{1}{\sqrt{2+\gamma^2}}(\vert 0 \rangle_A \vert 0\rangle_B+\gamma\vert 1 \rangle_A \vert 1\rangle_B+\vert 2 \rangle_A \vert 2\rangle_B),
\end{equation}
where we associated the idler and signal photon with subsystem $A$ and $B$. For each of the two subsystems individual measurement settings $a,b\in \{1,2\}$ and measurement outcomes $A_a,B_b=0,\ldots d-1$ are assigned. According to CGLMP \cite{collins2002}, the generalized Bell parameter in $d$-dimensions is then defined as
\begin{eqnarray}
I_{d}&\nonumber \equiv \sum\limits_{k=0}^{[d/2]-1}\left(1-\dfrac{2k}{d-1}\right) &\lbrace +[P(A_1=B_1+k)+P(B_1=A_2+k+1)\\
\nonumber&& +P(A_2=B_2+k)+P(B_2=A_1+k)]\\
\nonumber&& -[P(A_1=B_1-k-1)+P(B_1=A_2-k)\\
&& +P(A_2=B_2-k-1)+P(B_2=A_1-k-1)]\rbrace 
\end{eqnarray}
and explicitly reads  
\begin{eqnarray}
I_{2}=I_{3}&\nonumber = &+P(A_1=B_1)+P(B_1=A_2+1)+P(A_2=B_2)\\ 
\nonumber&& +P(B_2=A_1)-P(A_1=B_1-1)-P(B_1=A_2)\\ 
&& -P(A_2=B_2-1)-P(B_2=A_1-1)
\end{eqnarray}
for $d=2,3$. For all $d\geq2$, the inequality $I_d\leq2$ holds for local realistic theories. The probability that the outcome $A_a$ differs from $B_b$ by $k$ modulo $d$ is thus given by
\begin{equation}
P(A_a=B_b+k)\equiv\sum_{j=0}^{d-1}P(A_a=j,B_b=(j+k)\,\text{mod}\,d).
\end{equation}
Analogue, we have
\begin{equation}
P(B_b=A_a+k)\equiv\sum_{j=0}^{d-1}P(A_a=(j+k)\,\text{mod}\,d,B_b=j).
\end{equation}
The Bell parameter is described by a combination of projective measurements on the states $\vert\chi_{m,n}\rangle=\vert m\rangle_A^{a}\vert n\rangle_B^{b}$ defined by
\begin{equation}\label{eq:basisA}
\vert m\rangle_A^{a}=\frac{1}{\sqrt{d}}\sum_{j=0}^{d-1}\exp\left(i\frac{2\pi}{d}j(m+\alpha_a)\right)\vert j\rangle_A^{a},
\end{equation}
\begin{equation}\label{eq:basisB}
\vert n\rangle_B^{b}=\frac{1}{\sqrt{d}}\sum_{j=0}^{d-1}\exp\left(i\frac{2\pi}{d}j(-n+\beta_b)\right)\vert j\rangle_B^{b},
\end{equation}
with $m,n=0,\ldots, d-1$ and a specific choice of detection settings $\alpha_1=0, \alpha_2=1/2, \beta_1=1/4,$ and $\beta_2=-1/4$ for maximally entangled states  \cite{collins2002}. Allowing for the states in Eq.~(\ref{eq:gamma_2}) and Eq.~(\ref{eq:gamma_3}), the individual joint-probabilities become a function of $\gamma$. In accordance with Eq.~(\ref{eq:projection}), a measured coincidence signal is given by 
\begin{equation}
P(A_a=m,B_b=n,\gamma)\propto\left\vert\langle\chi_{m,n}\vert\psi(\gamma)\rangle^{(d)}\right\vert^2 
\end{equation}
with the projective states $\vert\chi_{m,n}\rangle=\vert m\rangle_A^{a}\vert n\rangle_B^{b}$ of Eqs.~(\ref{eq:basisA},\ref{eq:basisB}). For a specific detection setting $a,b$ each coincidence measurement has to be normalized according to $P(A_a=m,B_b=n,\gamma)/\mathcal{N}_{a,b,\gamma}$ with $\mathcal{N}_{a,b,\gamma}=\sum_{m,n=0}^{d-1}P(A_a=m,B_b=n,\gamma)$. Starting from a maximally entangled state through Procrustean filtering, the reduced entanglement, i.e.~$\gamma<1$, is obtained by decreasing the transmission amplitudes of the bins associated with  $\vert 1 \rangle_A \vert 1\rangle_B$ using the SLM.

It has been shown \cite{acin2002} for qutrits that the Bell inequality $I_3\leq 2$ is maximally violated for $\gamma_{max}=(\sqrt{11}-\sqrt{3})/2 \approx 0.792$.
\begin{figure}[ht!]
\epsfig{file=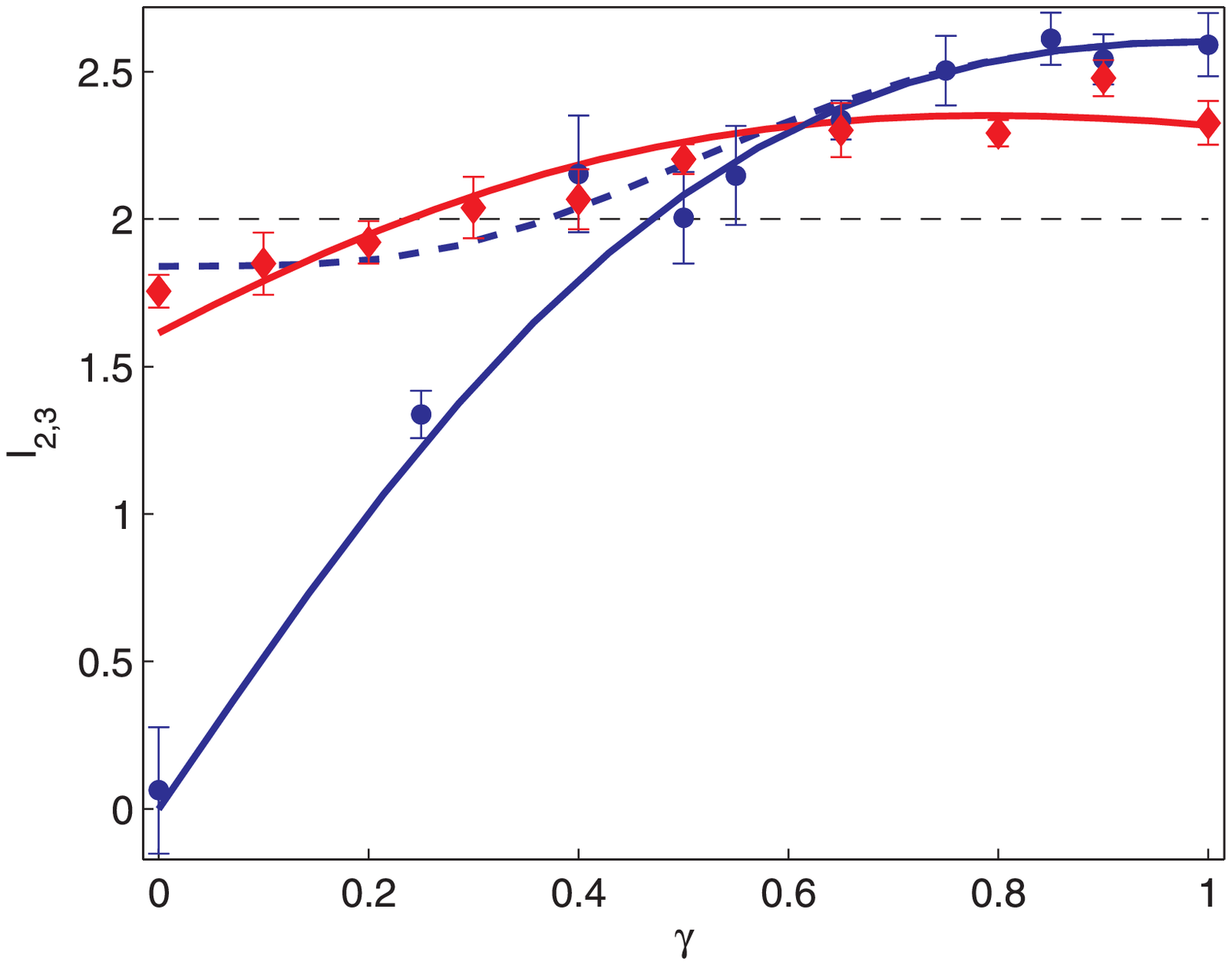,width=0.8\linewidth}
\caption{\label{fig:Igamma} $I^{exp}_2$ (blue dots) and $I^{exp}_3$ (red diamonds) show the experimental values of the Bell parameter in dependence of the entanglement parameter $\gamma$. The $2\sigma$-errors are calculated assuming Poisson statistics on background-subtracted coincidence counts. Straight lines show the theoretically predicted Bell parameters $I_2(\gamma)$ (blue),  $I_2(\gamma)$ using the Horodecki theorem (dashed blue), and $I_3(\gamma)$ (red). The curves are scaled with their corresponding mixing parameter. The dashed black line indicates the local variable limit.}
\end{figure}
We measured the Bell parameters $I_2^{exp}$ and $I_3^{exp}$ for qubits and qutrits as a function of $\gamma$ (Fig.~\ref{fig:Igamma}). The experiment reveals a higher sensitivity to $\gamma$ of the Bell parameter for qubits compared to qutrits, which is in accordance with theoretical predictions. The theoretical curves are scaled to the experimental data using the symmetric noise model $\hat{\rho}_{d}^{sn}(\gamma)=\lambda_d\vert\psi(\gamma)\rangle^{(d)}\,^{(d)}\langle\psi(\gamma)\vert+(1-\lambda_d)\mathbbm{1}_{d^2}/{d^2}$
in which deviations from a pure state due to white noise are quantified by a mixing parameter $\lambda_d$ and $\mathbbm{1}_{d^2}$ denotes the $d^2$-dimensional identity operator. The value of the Bell parameter for $\hat{\rho}_{d}^{sn}(\gamma)$ then scales as \mbox{$I_{d}(\hat{\rho}_{d}^{sn}(\gamma))=\lambda_d I_{d}(\gamma)$}. We experimentally determine the mixing parameters to be $\lambda^{exp}_2=0.920\pm0.013$ and $ \lambda^{exp}_3=0.807\pm0.008$ where the 2$\sigma
$-errors are calculated assuming Poisson statistics.
Note, that  the specific detection settings are not optimal for $d=2$. In Figure \ref{fig:Igamma}, we therefore additionally depict values of the Bell parameter for optimal settings given by Horodecki's theorem \cite{horodecki1995}. Similar to the measured $I_2(\gamma)$ and in contrast to $I_3(\gamma)$, the Horodecki curve is monotonically decreasing for $\gamma<1$. In the $d=3$ case the optimal settings were only determined for $\gamma=1$ and $\gamma_{max}$ \cite{acin2002}. Recent numerical studies in \cite{gruca2012} report an even stronger violation of local realism by two entangled qutrits provided that a more general measurement bases is applied than the settings used in the CGLMP inequalities.\\

\section{Conclusion}
By exploiting the flexibility of a SLM, we have been able to reconstruct maximally entangled qu$d$it states up to $d=4$ through quantum state tomography. We moreover demonstrated the existence of frequency entanglement by measuring a CGLMP Bell parameter above the local variable limit for maximally and certain non-maximally entangled qubit and qutrit states. In our actual experiment, the available dimensions to encode qu$d$its in the frequency domain are currently constrained by the finite resolution of the optical setup and the pixel size of the SLM. The spectral resolution can be improved from 9 nm to 0.2 nm by replacing prisms with gratings such that the accessible dimension becomes then only limited by the number of pixel of the shaping device. This modification would result in a seven-fold increase in dimension. The here discussed method allows for an easy access to the spectral components of the photons and, further, a broad class of transfer functions can be implemented on the SLM. Therefore, other qu$d$it encoding schemes like time-bins and realizations based on Schmidt modes can be carried out. In addition, with shaped frequency entangled qu$d$its, it is possible to investigate recently proposed theoretical results: Hilbert space dimensions can be probed using new dimension witnesses \cite{brunner2008} and extended Bell-tests with generalized entangled qutrit states are within reach \cite{gruca2012}.\\
Ultimately, the dimension of the Hilbert space is limited by the bandwidth of the pump laser and can reach values as large as a few millions. This provides a vast potential of encoding high-dimensional qu$d$its for quantum information and communication technologies.\\

\textbf{Acknowledgments}\\
This research was supported by the grant PP00P2 133596 and by NCCR MUST, both funded by the Swiss National Science Foundation.

\end{document}